\documentclass[rnote,traditabstract]{aa}

\usepackage{graphicx}
\usepackage{txfonts}
\usepackage{rotating}
\usepackage{natbib}

\def\Msun{M$_\odot$}
\def\Mbh{${\cal M}_{\rm BH}$}

\def\Edd{$L/L_{\rm Edd}$}
\def\lLl{$\lambda L_{\lambda}$}

\def\Civ{C\,{\sc iv}}
\def\Mgii{Mg\,{\sc ii}}
\def\Feii{Fe\,{\sc ii}}

\def\Hi{H\,{\sc i}}
\def\Oii{[O\,{\sc ii}]}
\def\Oiii{[O\,{\sc iii}]}
\def\Neiii{[Ne\,{\sc iii}]}

\def\Hg{H$\gamma$}
\def\lsim{\mathrel{\rlap{\lower 3pt \hbox{$\sim$}} \raise 2.0pt \hbox{$<$}}}
\def\gsim{\mathrel{\rlap{\lower 3pt \hbox{$\sim$}} \raise 2.0pt \hbox{$>$}}}

\def\s0927{S0927}

\begin{document}

\title{
 A quasar companion to the puzzling quasar SDSS J0927+2943\thanks{
 Based on observations collected at Asiago observatory.
 }
}
   \author{R. Decarli\inst{1,2}
          \and
          R. Falomo\inst{3} \and A. Treves\inst{1,4} \and M. Barattini\inst{1}
          }

   \institute{Universit\`{a} degli Studi dell'Insubria, via Valleggio 11,
             22100 Como, Italy
         \and
             MPIA, K\"{o}nigstuhl 17, 69117, Heidelberg, Germany
              \email{decarli@mpia-hd.mpg.de}
         \and
         INAF - Osservatorio Astronomico di Padova, Vicolo dell'Osservatorio 5,
             35122, Padova, Italy
         \and
         Istituto Naz. Fis. Nucleare - Universit\`a di Milano-Bicocca, Piazza della Scienza 3, 20126 Milano, Italy\\
             }

   \date{ }


\label{firstpage}

\abstract{
We report the discovery of a quasar close to SDSS J0927+2943
($z = 0.713$), which is a massive binary / recoiling black hole
candidate. The companion quasar is at a projected distance of
125 $h_{70}^{-1}$ kpc and exhibits a radial velocity difference of
$\sim$ 1400 km/s with respect to the known quasar.
We discuss the nature of this peculiar quasar
pair and the properties of its environment. We propose that
the overall system is caught in the process of ongoing
structure formation.
}


\maketitle

\section{Introduction}

Quasar pairs are usually divided in: {\it i-} physical
pairs, in which the two quasars have similar redshift and belong to the
same cosmological structure \citep[e.g.,][]{foreman08}; {\it ii-}
gravitational lenses, where the light of a single quasar is split into
two or more images due to the light bending of an intervening massive
object \citep[e.g.][]{wittman00,chieregato07}; {\it iii-} apparent pairs,
resulting from chance projected associations. Each of the three classes
has great importance for probing the galactic halos of quasar
host galaxies (column density, metal and dust abundances, ionization, etc),
the distribution of matter from galactic to super-cluster scales,
and the role of galaxy interactions in triggering nuclear activity.
For instance, \citet{kirkman08} and \citet{gallerani08} used absorption
features in the spectra of apparent quasar pairs to study the ionization
properties of the gas in the halos of quasar hosts and to constrain the
duty cycle of the nuclear activity.
\citet{zhdanov01}, \citet{myers07} and \citet{hennawi06,hennawi09} used
close, physical quasar pairs to show that the quasar correlation function
gets progressively steeper at sub-megaparsec scales, where gravitational
interactions among galaxies become stronger.

Quasar pairs are very rare: Only few tens of quasar pairs are known
with sub-arcmin separation \citep{vcv06,hennawi06}. In a previous
work \citep{dtf09}, we estimated that, given a quasar, the chance
probability of finding a companion quasar with $m_b\lsim20$ within
$\approx10''$ is $4\times10^{-4}$ \citep[assuming the quasar
number density by][]{croom04}. On the other hand, quasars belonging
to the same physical structure show significant clustering
\citep[see, e.g.,][]{coil07,bonoli09,shen09}. At very small angular
separations, the surface density of quasars within dense
environments is up to 3 orders of magnitude higher than in the field
\citep[see, for instance,][]{hennawi06,wrobel09}. Wide-field
surveys, such as the Sloan Digital Sky Survey \citep[SDSS; see]
[]{york00} collected spectra of $\sim100,000$ quasars, but missed
most of close pairs because of the finite physical dimension of the
spectroscopic fibers. On the other hand, the enormous, multi-band
imaging database of the SDSS allows the search for quasar candidates
starting from their photometry. For instance, \citet{hennawi09}
employed colour-selection techniques and spectroscopic follow-up
observations to discover 24 new physical quasar pairs at $3.0 < z <
4.5$.

In this framework, we started a programme to search for low-redshift
quasar pairs with small separations, starting from the SDSS photometric
dataset \citep[Decarli et al., in preparation; see also][]{d1536,dtf09}.
Low-redshift quasars have optical Spectral Energy Distributions (SEDs)
similar to those of blue stars \citep[see, e.g.,][]{richards02}.
Starting from the $z<1$ quasars in the spectroscopic catalogue by
\citet{schneider07}, we searched for quasar companions in the SDSS
photometric database with projected separation below 200 $h_{70}^{-1}$
kpc (at the redshift of the known quasar). Quasar companion candidates
are selected on the basis of their colours, independently of the properties
of the reference quasars. We limited our study to Galactic
latitude $b>45^\circ$ to minimize contaminations from stars.
About one hundred quasar pairs are selected in this way. For twenty
of them the companion object was also detected by the GALaxy
Evolution EXplorer \citep[GALEX;][]{siegmund04} in the {\it FUV} and
{\it NUV} bands. For these systems, we designed a plan of follow-up
spectroscopic observations at the Asiago Telescope, in order to
confirm the quasar classification and to study the properties of the
quasar pair.

Within this line of research, which at the moment had covered only few objects,
here we present the discovery of a companion quasar of the peculiar
quasar \object{SDSS J092712.65+294344.0} (hereafter, \s0927), a promising massive black
hole binary / recoiling black hole candidate (see section
\ref{sec_s0927}).
The present discovery of a second quasar with $\Delta v\sim1400$ km/s
along the line of sight and projected separation $\sim125$ $h_{70}^{-1}$
kpc adds new, unexpected elements for our comprehension of
\s0927.

In section \ref{sec_obs} we present our observations and the
data reduction. Results are given in section \ref{sec_res}.
In section \ref{sec_disc} we discuss the nature of this quasar pair,
focussing in particular on its environment.
Throughout the paper, we adopt a concordance cosmology
with $H_0=70$ km/s/Mpc, $\Omega_m=0.3$, $\Omega_\Lambda=0.7$.

\section{The unusual quasar SDSS J0927+2943}\label{sec_s0927}

\s0927{} (RA(J2000): $09^{\rm h}27^{\rm m}12^{\rm s}.6$, Dec(J2000):
$+29^{\rm d}43^{\rm m}44^{\rm s}$; $u$=$18.69$, $g$=$18.42$,
$r$=$18.40$, $i$=$18.40$, $z$=$18.34$) is a puzzling quasar,
discovered by \citet{komossa08} out of the enormous SDSS
spectroscopic database. It shows two sets of narrow emission lines
at different redshifts (hereafter, the `blue' system at $z_{\rm
b}=0.698$ and the `rest-frame' system at $z_{\rm rf}=0.713$; $\Delta
v\approx2650$ km/s), and a set of broad emission lines at $z_{\rm
b}$. The existence of another set of emission lines at
$z\approx0.703$ was claimed by \citet{shields09}, mainly based on
the detection of an emission line at 8526 \AA{}, identified with the
\Oiii{}$_{\lambda 5007}$. We note however that this line is also
consistent with the \Feii{}$_{\lambda 5014}$ at $z=z_{\rm b}$, a
common feature in quasar spectra.

Three scenarios have been proposed to explain the velocity
difference between $z_{\rm b}$ and $z_{\rm rf}$:
\citet{komossa08} suggested that the active black hole (BH)
in this object is recoiling as the result of the coalescence of
two massive BHs; \citet{dotti09} and \citet{bogdanovic09}
proposed that the active black hole is part of a massive BH
binary with sub-parsec separation; finally \citet{heckman09} proposed
that the velocity difference in the two systems in \s0927{} is due
to the ongoing merger event between a massive galaxy, hosting the quasar,
and a satellite (responsible of the emission of narrow lines at
$z_{\rm rf}$). This last scenario relies on the assumption
of a close alignment between the two galaxies, which is statistically
acceptable only if \s0927{} resides in a rich cluster. Furthermore, the
potential well of a rich galaxy cluster would explain the velocity
difference between the two emission line systems in \s0927{}. On the
other hand, \citet{drd09} showed that the field of this object is not as
rich as the cluster scenario requires, the number of possible galaxy
candidates being consistent with a rich group in the best case.

\section[]{Observations and data reduction}\label{sec_obs}

\begin{figure}
\begin{center}
\includegraphics[width=0.49\textwidth]{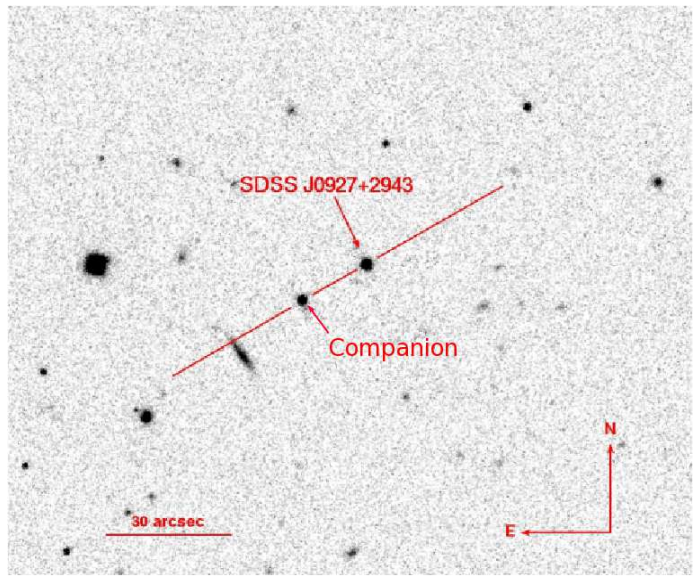}\\
\caption{The field of \s0927{} as imaged in the $r$ band
from the SDSS. The slit orientation adopted in our new
observation is also plotted.
}\label{fig_ima}
\end{center}
\end{figure}

The optical spectrum of \s0927{} was collected with the $1.82$m Cima
Ekar telescope at the Asiago Observatory on January, 4, 2009. The
Asiago Faint Object Spectrograph Camera was mounted in long-slit
spectroscopy configuration with grism \#4, yielding a spectral
resolutions of $R\sim 300$ ($2.10"$ slit) in the spectral range
$3500$--$7800$ \AA{} ($\Delta\lambda/$pixel = $4.24$ \AA). At
$\lambda \approx 5000$ \AA{} the spectral instrumental resolution is
$\sim 17$ \AA. The slit was oriented with Position
Angle=$30.5^\circ$, so that the spectrum of the blue source placed
$17.5''$ South-West of the main target could be simultaneously
acquired (see Figure \ref{fig_ima}). The total integration time was
100 minutes.

The standard IRAF\footnote{IRAF is distributed by the National
Optical Astronomy Observatories, which are operated by the
Association of Universities for Research in Astronomy, Inc., under
cooperative agreement with the National Science Foundation.}
procedure was adopted in the data reduction. The \verb|ccdred|
package was employed to perform bias subtraction, flat field
correction, frame alignment and image combination. Cosmic rays were
eliminated with the \verb|cosmicrays| task in the \verb|crutils|
package. The spectra extraction, the background subtraction and the
calibrations both in wavelength and in flux were performed with
\verb|doslit| task in \verb|specred| package, using a Hg-Cd lamps
and the spectrophotometric standard star Feige 34 as reference.
Wavelength calibration residuals are around $0.2$ \AA{} (sub-pixel).
Absolute calibration of spectra was optimized through the photometry
of field stars, by comparing corollary imaging with Johnson's R
filters to the magnitudes published in the U.S. Naval Observatory
catalogue. The uncertainty in the flux calibration is $0.1$ mag.
Galactic extinction was accounted for according to the Galactic
\Hi{} maps by \citet{schlegel98} and assuming $R_V = 3.1$.

\begin{figure}
\begin{center}
\includegraphics[width=0.49\textwidth]{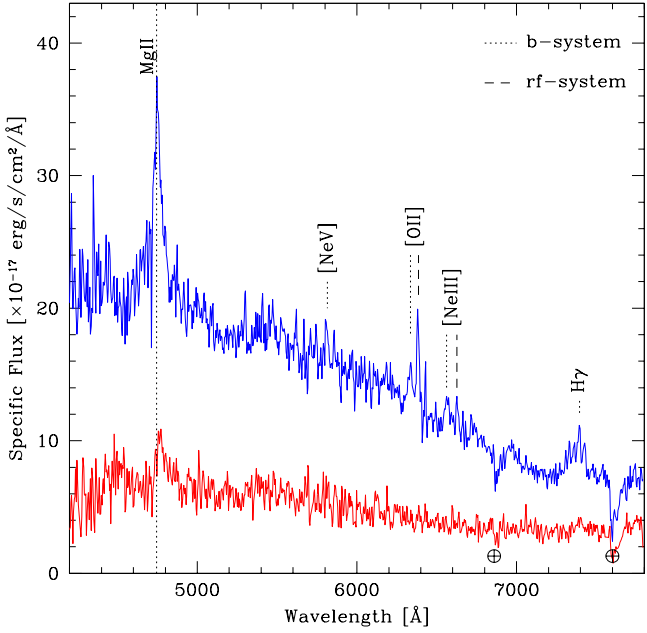}\\
\caption{Our new spectra of \s0927{} and its (fainter) companion.
Main emission lines in the spectrum of \s0927{} are labelled
\citep[see also][]{komossa08}. The Earth symbols mark relevant
atmospheric absorption features. Note the detection of the \Mgii{}
line in the spectrum of quasar B and a tentative detection of
\Hg{}, confirming the line identification.
The Signal-to-Noise ratios per pixel in the two spectra are
$\sim20$ and 8.
}\label{fig_spectra}
\end{center}
\end{figure}

The spectra of the two sources are shown in Figure \ref{fig_spectra}.

\begin{table}
\begin{center}
\caption{Properties of the detected emission lines. (1) Line. (2) Peak
wavelength from the line fit. (3) Redshift corresponding to
$\lambda_{\rm peak}$. (4) Fitted FWHM in the observed frame. (5) Line
Equivalent Width in the observed frame.
} \label{tab_lines}
\begin{tabular}{ccccc}
   \hline
   Line          & $\lambda_{\rm peak}$       & $z$   & FWHM  & E.W.      \\
                 &  [\AA]                     &       & [\AA] & [\AA]     \\
   (1)           &  (2)                       &  (3)  & (4)   & (5)       \\
   \hline
   \multicolumn{5}{l}{{\it \s0927{}, blue system}} \\
   \Mgii{} (broad)   & $4750\pm5$             & 0.697 & $ 68\pm17$ & $85\pm11$ \\
   \Oii{}            & $6335\pm5$             & 0.699 & $ 18\pm10$ & $7\pm4$   \\
   \Neiii{}          & $6565\pm6$             & 0.697 & $ 30\pm8$  & $6\pm3$   \\
   H$\gamma$ (broad) & $7379\pm9$             & 0.700 & $100\pm20$ & $39\pm8$  \\
   \hline
   \multicolumn{5}{l}{{\it \s0927{}, rest frame system}} \\
   \Oii{}            & $6382\pm1$             & 0.712 &  $17\pm6$  & $13\pm2$  \\
   \Neiii{}          & $6628\pm3$             & 0.713 &  $17\pm10$ & $4\pm3$   \\
   \hline
   \multicolumn{5}{l}{{\it Companion quasar}} \\
   \Mgii{} (broad)   & $4772\pm8$             & 0.706 &  $55\pm20$ & $45\pm20$ \\
   H$\gamma$ (broad) & $7395\pm15$            & 0.704 &  $30\pm20$ & $22\pm15$ \\
   \hline
   \end{tabular}
   \end{center}
\end{table}

\section[]{Results}\label{sec_res}

\begin{figure}
\begin{center}
\includegraphics[width=0.49\textwidth]{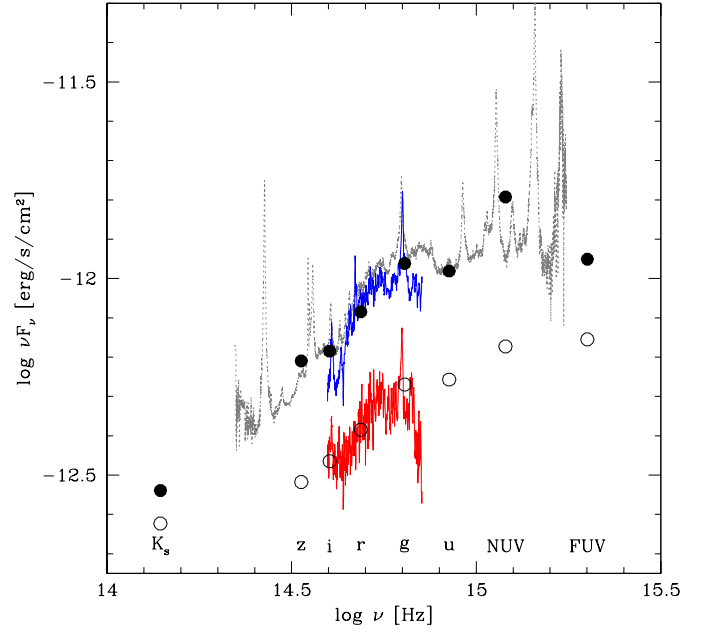}\\
\caption{The Spectra Energy Distributions of \s0927{} and its companion.
Observed spectra are plotted as solid lines, while photometry data
are reported as circles. The quasar composite spectrum derived in
\citet{paperI} is also shown for comparison (dotted line). {\it FUV}
and {\it NUV} fluxes derived from the GALEX archives. The magnitudes
in the optical bands are from the SDSS. The $K_s$ flux is
from \citet{drd09}. The SEDs of the two quasars are remarkably similar.
}\label{fig_sed}
\end{center}
\end{figure}

The blue object selected as a quasar candidate $17.5''$ far from
\s0927{} (\object{SDSS J092713.82+294335.5}; RA(J2000): $09^{\rm h} 27^{\rm
m} 13^{s}.8$, Dec(J2000): $+29^{\rm d} 43^{\rm m} 35^{\rm s}$; 
$u$=$19.38$, $g$=$19.18$, $r$=$19.14$, $i$=$19.01$, $z$=$19.13$)
appears in the USNO-A2 catalogue \citep{monet98}. The GALEX All-Sky
Survey provides estimates of its UV magnitudes ({\it FUV} = $19.94
\pm 0.19$, {\it NUV} = $19.22 \pm 0.11$). The blue optical colours
and the bright UV emission are consistent with the typical Spectral
Energy Distribution (SED) of a quasar \citep[see Figure
\ref{fig_sed} and][]{drd09}. The spectrum shows a broad emission
line at $4772$ \AA{} and another possible emission line at $7395$
\AA{} (see Figure \ref{fig_spectra} and table \ref{tab_lines}). We
identify the former with the \Mgii{} and the latter with \Hg{},
yielding $z=0.705$. Other identifications of the 4772 \AA{} line
(e.g., with \Civ{}) are ruled out by the lack of any correspondence
with other expected bright UV lines, e.g., C{\sc iii}], Si{\sc iv}
and Ly$\alpha$.

No obvious detection of the \Oii{} doublet
is reported. We fitted the profiles of these lines and
of the main lines in the spectrum of \s0927{} with gaussian profiles
\citep[a superposition of 2 gaussian is used for the \Mgii{} line; see
the discussion on the fit of quasar broad emission lines in][]{paperI}.
Table \ref{tab_lines} reports the best fit parameters for the peak
wavelength and corresponding $z$, for the line width (as measured in the
observed frame) and for the line equivalent width. Referring to standard
techniques for single-epoch spectra of quasar \citep[e.g.,][]{paperI}, we
infer the BH mass of the newly discovered quasar from the continuum
luminosity and the \Mgii{} line width: \Mbh{} $\sim1.4\times10^{8}$
\Msun{}. The uncertainty on \Mbh{} is of a factor $\sim2$, dominated
by the dispersion on the adopted broad line region radius--luminosity
relation.
The corresponding Eddington ratio is \Edd{}=$0.22$, assuming the
bolometric correction $L$/\lLl{}(3000 \AA)=$5.15$
\citep{richards02}.

\section{Discussion and conclusions}\label{sec_disc}

The observations of the quasar pair presented here reveal a puzzling
nature. The probability that the pair is due to a chance superposition
is very small: Following \citet{dtf09}, we estimate that the probability
to find a companion quasar
with $m_b<20$ and angular separation $<20''$ {\it at any
redshift} is $\sim 10^{-3}$. If we consider pairs
with redshift difference $< 1,500$ km/s, the probability drops by
two orders of magnitude. Hence, at most one quasar pair with
properties similar to those observed in the case of \s0927{} is expected
among the $\sim 77,000$ quasars in the SDSS catalogue by \citet{schneider07}.
Therefore the chance superposition is very unlikely.

Alternatively, the two quasars may belong to a common physical structure,
as suggested by their velocity difference ($\approx-1400$ km/s with
respect to $z_{\rm rf}$, $\approx 1200$ km/s with respect to $z_{\rm b}$).
Quasars show some evidence of clustering in a way
that roughly resembles what observed in quiescent galaxies
\citep{sochting02,coil07,shen09}. Different environments are
observed depending on the radio loudness \citep{barr03,sochting04} or on
other AGN properties \citep[e.g.,][]{hickox09}. \citet{boris07} analyzed
the environment of 4 quasar pairs at $z\sim1$, and found evidence of
rich galaxy environment in 3 of them, while one pair (QP 0114-3140)
appears isolated. \citet{djorgovski07} discovered a quasar triplet
at $z=2.076$, with relative velocities of few hundrends km/s. The presence
of many galaxies in the field around the triplet suggests that
its environment may be particularly dense.

In the case of the quasar pair of \s0927{}, if we assume that the two
objects are gravitationally bound, their velocity difference and projected
separation imply a dynamical mass of $\sim10^{14}$ \Msun{} (depending on the
de-projected separation and the direction of the velocity vector).
\citet{drd09} reported that the number of galaxies around this system is
consistent with the presence of at most a moderately rich galaxy group.
Unless extreme Mass-to-Light ratios are invoked, this would suggest
that the system is not virialized. This quasar pair might thus unveil the
occurrence of an ongoing structure formation, similarly to the
low-redshift cases of the Blue Infalling Group \citep{gavazzi03,cortese06},
the Stephan's Quintet \citep{sulentic01} or the Cartwheel's system
\citep{taylor84,wolter99}.
The main difference among these low-$z$ counter-parts and \s0927{} is
that none of the former examples shows quasar-like nuclear activity
in any galaxies (but note that NGC 7319 in the Stephan's Quintet
is a Seyfert 2). We remark here that a systematic, spectroscopic study
of the galaxies observed in the field of \s0927{} is mandatory to definitely
pin down the dynamical state of this system. Further, deep images
both in the optical and X-ray bands would also provide new contraints on
the mass and structure of the galactic environment of this pair.

Our results provide additional clues supporting the presence of a galaxy
group surrounding \s0927{}. This kind of structure is the ideal
habitat for strong gravitational interactions to occur, which may trigger
quasar-like nuclear activity \citep{canalizo07,bennert08,bennert09}.
Alltogether, the available information about this system support a view
in which a recent or ongoing galaxy merger is present. On the other
hand, the projected distance and the relative velocity of the quasar pair
are too high to account per se for explaining the occurrence of two
emission line systems in \s0927{}, which appears independent of the
presence of a companion quasar. Deep, high-resolution
multi-band images and multi-object spectroscopy of the sources in this
field are required to probe the actual build-up of a galaxy group,
to search for signatures of gravitational interactions, and to pin down
the dynamics of the system.

\section*{Acknowledgments}
We thank the referee for his/her useful comments. R.D. thanks
Massimo Dotti for useful discussions on the nature of this system.
This work was based on observations collected at Asiago observatory.
This research has made use of the NASA/IPAC Extragalactic Database
(NED) which is operated by the Jet Propulsion Laboratory, California
Institute of Technology, under contract with the National
Aeronautics and Space Administration.

\label{lastpage}
\end{document}